# Model for Thermal Comfort and Energy Saving Based on Individual Sensation Estimation

Guillaume Lopez,* Takuya Aoki, Kizito Nkurikiyeyezu, and Anna Yokokubo

Aoyama Gakuin University, Graduate School of Science and Engineering,
5-10-1 Fuchinobe, Chuo-ku, Sagamihara, Kanagawa 252-2558, Japan

In office spaces, the ratio of energy consumption of air conditioning and lighting for maintaining the environment comfort is about 70%. On the other hand, many people claim being dissatisfied with the temperature of the air conditioning. Therefore, there is concern about work efficiency reduction caused by the current air conditioning control. In this research, we propose an automatic control system that improves both energy saving and thermal comfort of all indoor users by quantifying individual differences in thermal comfort from vital information, on the basis of which the optimal settings of both air conditioning and wearable systems that can directly heat and cool individuals are determined. Various environments were simulated with different room sizes, numbers of users in a room, and heating/cooling conditions. The simulation results demonstrated the efficiency of the proposed system for both energy saving and comfort maximization.

## 1. Introduction

In office spaces, air conditioning accounts for the majority of energy consumption. On the other hand, several surveys reveal that many people claim being dissatisfied with the thermal office environment, leading to a decrease in productivity.(1,2) Indeed, despite decades of research and the undeniable progress, the provision of thermal comfort in old buildings is still marred by problems in terms of both its quality and the amount of energy consumed.(3) Despite dedicated energy resources, thermal comfort in buildings remains unsatisfactory. A survey of buildings in the US, Canada, and Finland found that only 11% of the buildings achieved the recommended thermal comfort satisfaction rate.(4) This finding agrees with those of the International Facility Management Association (IFMA) that pinpointed a high level of dissatisfaction with thermal comfort in buildings.(5) Indeed, the rationale of thermal neutrality has been contested as a misleading concept since, in reality, people prefer non-neutral conditions.(6)

Most thermal provision systems are based on the Predicted Mean Vote (PMV)/Predicted Percentage of Dissatisfied (PPD) model, with which the isothermal settings for all building occupants can be prescribed.(7) Although it performs well when its assumptions are met, some of its premises are ignored in practice.(8)

*Corresponding author: e-mail: guillaume@it.aoyama.ac.jp

Also, in Japan, comfortable lifestyles and living environments realized owing to abundant electric power supplies are in consistent with energy consumption reduction policies. In the summer of 2005, the Government of Japan, through the Agency for Natural Resources and Energy, started the "cool biz" campaign, which recommends setting the room air conditioner temperature to 28 °C for energy-saving promotion,[9] instead of the usual average room temperature setting of 25 °C in about 16 thousand office buildings in the Tokyo metropolitan area. Most companies followed this guideline; however, several surveys showed an increase in thermal dissatisfaction rate and a reduction in productivity.[10,11] Moreover, this has been accompanied by an increase in heatstroke risk owing to not only the hot but also humid summer in Japan. According to the Japan Meteorological Agency, more than 70 summer days exceed a daily temperature of 30 °C and the average relative humidity is about 70%. When the air temperature is 30 °C and the relative humidity is 70%, the wet bulb globe temperature (WBGT, a measure of heat stress) is 29 °C, which corresponds to a risk category requiring strict vigilance according to heatstroke prevention regulations.

However, it has been shown that differences in thermal comfort between individuals are large with a maximum difference of about 7 °C in the preferred and comfortable temperature range,[12] and many studies target a thermal comfort range for 80% of the people, so it is difficult to realize a comfortable environment for all people.[13] Since it depends on personal sensation, in dynamic community spaces such as coworking or coliving spaces, it is by definition not possible to satisfy the thermal comfort of all the users of such community spaces with one single setting of the air conditioning system,[14] creating thermal dissatisfaction for everyone in the community. Thus, it is clear that a comfortable thermal environment in coworking spaces cannot be realized using only a global heating and cooling system (GHCS) such as an air conditioner.

On the other hand, many research studies on the effects of thermal stress deal with heat stress. Extreme temperatures or long exposure to less-than-ideal weather can change core body temperature and impact homeostatic control, i.e., the body's ability to maintain its own temperature. As the body works harder to maintain a healthy core temperature by reallocating resources such as water and energy, the brain is deprived of these same resources and one's ability to think declines.[15] Several research studies showed that both local cooling and warming affect variations in physiological indices.[16–18] Effective energy saving technology for the direct temperature conditioning of the human body has been proposed, and Peltier-element-based portable systems that can both cool and warm the neck or wrist have been developed.[19–21] Although the comfort sensation in one environment is different among individuals and it can be different for the same person under different conditions, the current system temperature control algorithm depends on only environmental conditions and not the wearer's.

In a previous work, we developed a model that can detect the thermal comfort state of a user from heart rate variability indices.[22] On the basis of this model, we propose an automatic control system that improves both the energy saving and thermal comfort of all indoor users not only by combining GHCS with an individual heating and cooling system (IHCS) that can directly heat and cool individuals but also by quantifying individual differences in thermal comfort on the basis of biological information to optimize the combination (Fig. 1).

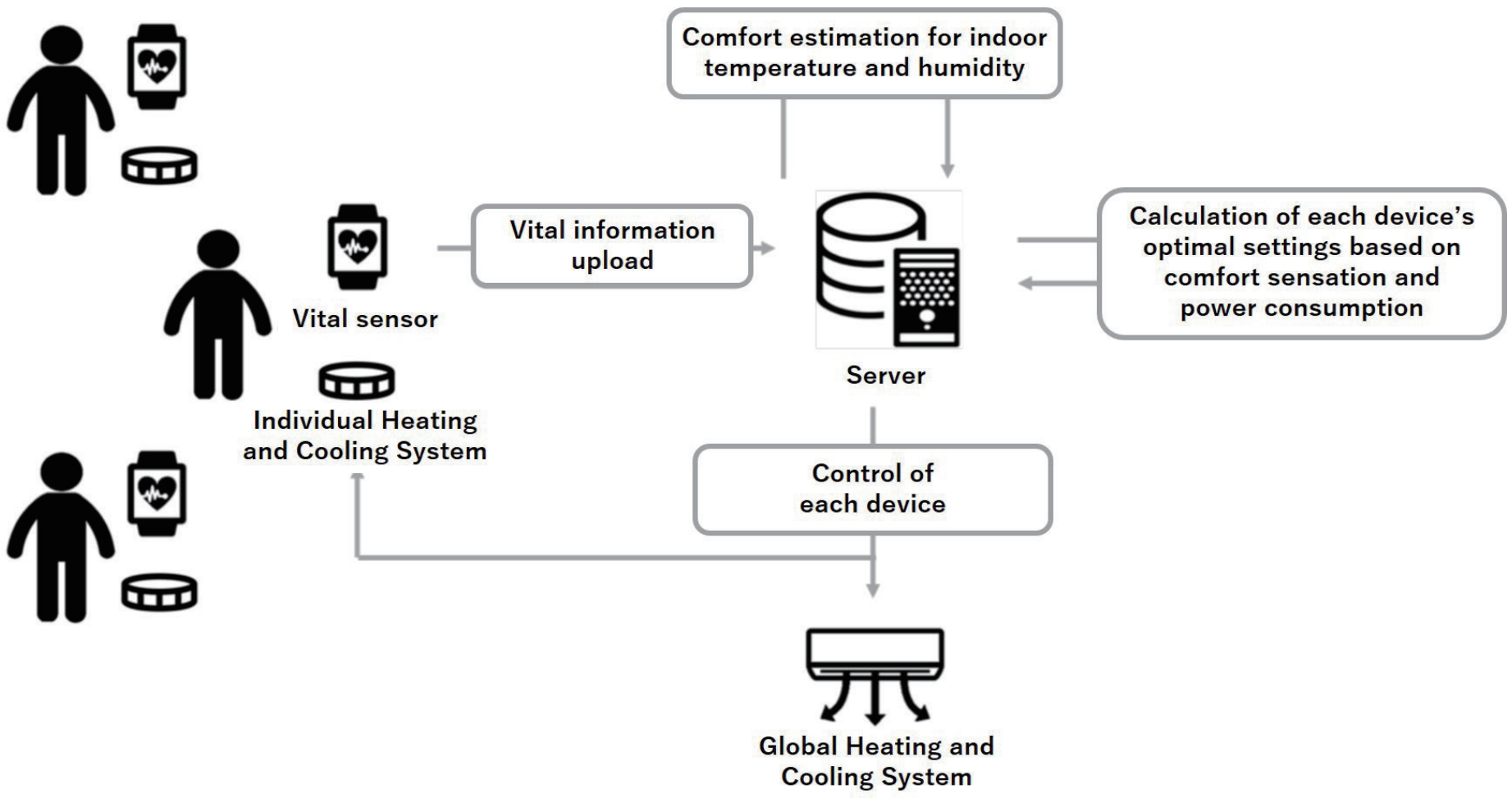

Fig. 1. Schematic of the proposed system.

## 2. Material and Methods

In this research, the following three points have been addressed in constructing the proposed system to automatically balance the thermal comfort for all users and energy saving in a coworking space:

1. modeling of user's current thermal comfort sensation, indoor temperature, humidity, and power consumption by GHCS activation, and user's thermal comfort sensation and power consumption by IHCS activation,
2. proposal of a control method that can improve both the thermal comfort sensation of all users and energy saving using the models, and
3. simulation of several thermal and office conditions and confirmation of the effect of the proposed method compared with the method using only GHCS or IHCS.

### 2.1 Conventional methods for thermal comfort sensation evaluation

Various indexes and evaluation methods have been defined so far for the evaluation of the thermal environment.[23] The typical ones are shown below.

- The discomfort index (DI) defines the environment's thermal discomfort and was shown to strongly correlate with thermal sensation.[24,25]
- PMV is an index of the heat balance between the human body and the environment, and correlates the deviation from the thermal average and thermal sensation, used in relation to the proportion of people who feel uncomfortable in an environment.[7]
- Standard new effective temperature (SET*) is an index developed by the American Society of Heating, Refrigerating and Air-Conditioning Engineers (ASHRAE) that can be used to compare different thermal environments.[13,26]

We can consider that by using these indices, the user's thermal sensation can be quantified as a parameter used in the proposed system. However, since these indices are consistent with the surrounding environmental conditions, and assuming the average of a specific sample group, it can be considered that the same model should not be applied for any given individual and situation as mentioned previously.[14] In our previous work, we demonstrated a model that can detect the thermal comfort state of a user from heart rate variability indices.[23] More recently, Kobiela *et al.* have obtained similar results under cold environment conditions.[27]

### 2.2 Modeling of individual thermal comfort sensation

To build an algorithm for achieving both energy saving and thermal comfort for all the users, it is necessary to model the power consumption and thermal comfort of each user, respectively. Conventional DI defines the environment thermal discomfort, which depends only on environmental conditions as described in Eq. (1).

$$DI = 0.81t + 0.01RH \times (0.99t - 14.3) + 46.3, \tag{1}$$

where $RH$ is the relative humidity and $t$ the dry bulb temperature.

A $DI$ between 65 and 70 is considered to be the most comfortable, and we define this range as the maximum thermal comfort range (MTCR). As a method of evaluating a user's thermal comfort, we propose the sensation of discomfort index (SDI). SDI is calculated by adding the individual difference of a user's thermal comfort and the correction of IHCS to the conventional DI, which defines the environment's thermal discomfort as

$$D_f(i, TD) = DI + D_p(i) + D_n(i, TD), \tag{2}$$

where $D_f(i,TD)$ is the SDI for user $i$ at the target discomfort index $TD$, $D_p(i)$ is the individual difference's correction of user $i$, $D_n(i,TD)$ is the correction by IHAC for user $i$, and $DI$ is the indoor discomfort index defined in Eq. (1).

To evaluate the power consumption, we use the total heat consumption (THC) calculated by adding GHCS and IHCS heat consumptions. The GHCS heat consumption is calculated by adding the amount of heat held by air, the amount of heat input from the outside, and the amount of heat generated by the users as

$$H_e = \alpha \,| H_v - H_f - H_p |, \tag{3}$$

where $H_e$ is the heat consumption of GHCS, $\alpha$ is the GHCS load, $H_v$ is the target temperature and humidity, $H_f$ is the heat from the outside, and $H_p$ is the amount of heat generated by the users.

### 2.3 Method for individual comfort sensation optimization and energy savings

On the basis of the individual thermal comfort estimation model proposed in Sect. 2.2, we propose a new control method that can improve both the thermal comfort sensation of all users and energy saving. We define the thermal comfort error as an absolute value of the difference between the user's SDI and the nearest boundary of the MTCR. Since the sum of error of all users ($E_S$) and THC can be uniquely determined using the target discomfort index (TDI) in the room, we defined the optimum setting to be the conditions that minimize both $E_S$ and THC as described in Fig. 2. Therefore, we examined the changes in $E_S$ and THC as functions of SDI. The results showed that the optimum setting is that at which the SDI has the smallest absolute difference from the current DI in the room within the MTCR. Accordingly, we propose using these settings the optimal control target.

## 3. Results

We conducted a simulation under the following three conditions: using only GHCS, using only IHCS, and using the proposed system that optimally combines GHCS and IHCS.

### 3.1 Simulation parameters and settings

The simulation parameters were set as follows.

- Each system was operating for 1 h.
- The standard volume of the room for the number of users was set as shown in Table 1.
- The outside temperature and humidity were respectively set to 30 °C and 60% for cooling and 12 °C and 60% for heating.

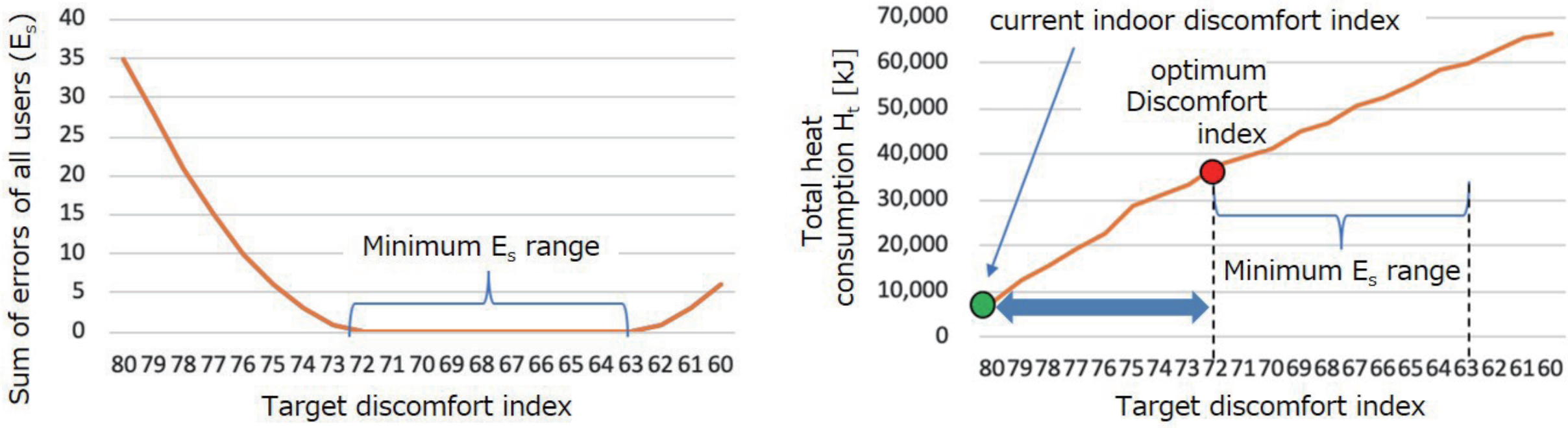

(a) (b)

Fig. 2. (Color online) Method of detecting (a) the optimum TDI from the changes in $E_S$ and (b) TDI from environment discomfort index.

Table 1
Standard volume of the room per maximum number of users.[29]

| Nb users | Room vol. ($m^3$) |
|---|---|
| 5 | 172.3 |
| 10 | 344.5 |
| 15 | 516.8 |
| 20 | 689.0 |
| 25 | 861.3 |

- We defined a correction model of SDI for individual differences in thermal comfort such that it can vary at seven levels from −3 to +3 compared with DI. A survey showed that the individual differences in the temperature preference of 29 persons were large, i.e., a difference of 7.2 °C between the maximum and minimum temperatures.(12) The individual differences were simulated according to a normal distribution on the scale from −3 to +3 (Fig. 3).
- We defined a correction model of SDI for the use of an IHCS, on the basis of the correction function of the IHCS prototype developed by the Advanced Institute of Wearable Environmental Information Networks.(28) In this study on 24 persons in a room at the same humidity (60%), when the temperature gradually increases, the average temperatures at which persons start feeling discomfort were 27.3 °C when wearing the IHCS and 31 °C when not. Since this is a difference of 5.2 in DI, the IHCS was assumed to be able to correct the DI up to 5. Therefore, when the IHCS is used, its effect was simulated by correcting the DI in 11 steps from −5 to +5 (Fig. 4)

### 3.2 Evaluation of efficiency of proposed control method by simulation

The results shown in Fig. 5 revealed that the power consumption reduction ratio when using the proposed system compared with that when using only GHCS increases for a larger indoor volume and fewer indoor users (30% maximum in cooling and 10% maximum in heating).

Concerning comfort optimization, as shown in Fig. 6, in the case of using only GHCS or IHCS, the sum of thermal comfort errors of all users ($E_S$) increases proportionally with the number of users. On the other hand, in the case of using the proposed system, which optimally combines GHCS and IHCS, $E_S$ is kept to 0 all the time, indicating that all users are thermally comfortable.

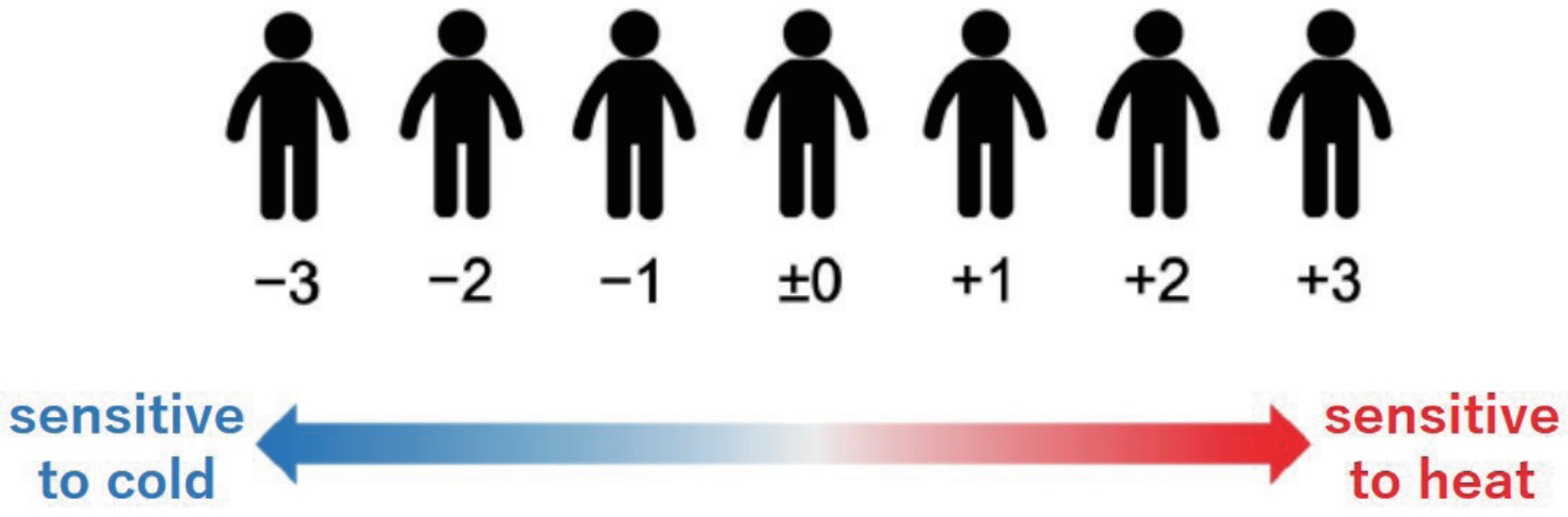

Fig. 3. (Color online) Correction of SDI by individual differences.

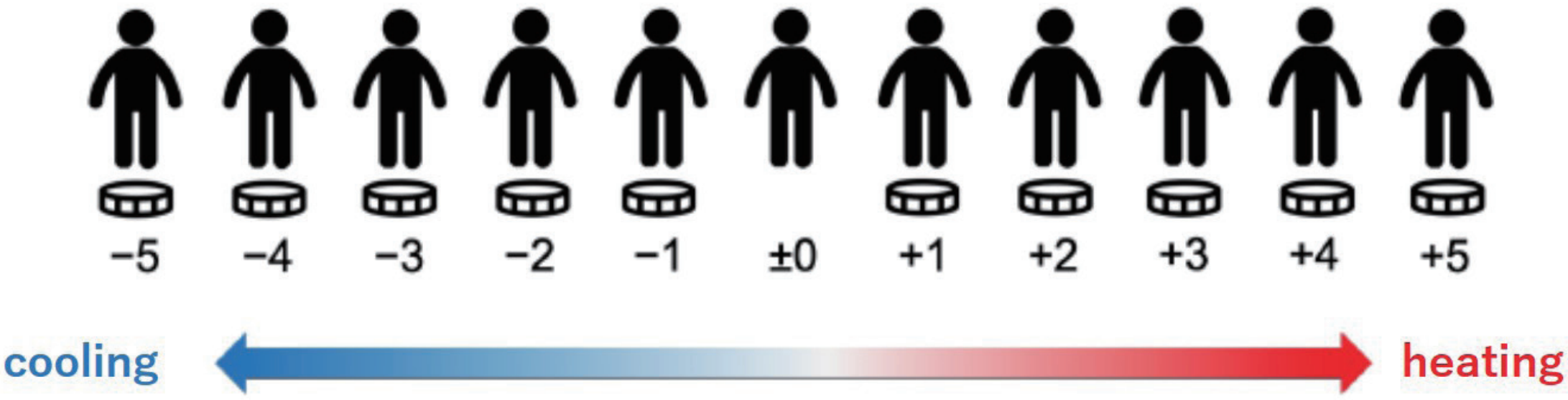

Fig. 4. (Color online) Correction of SDI by IHCS.

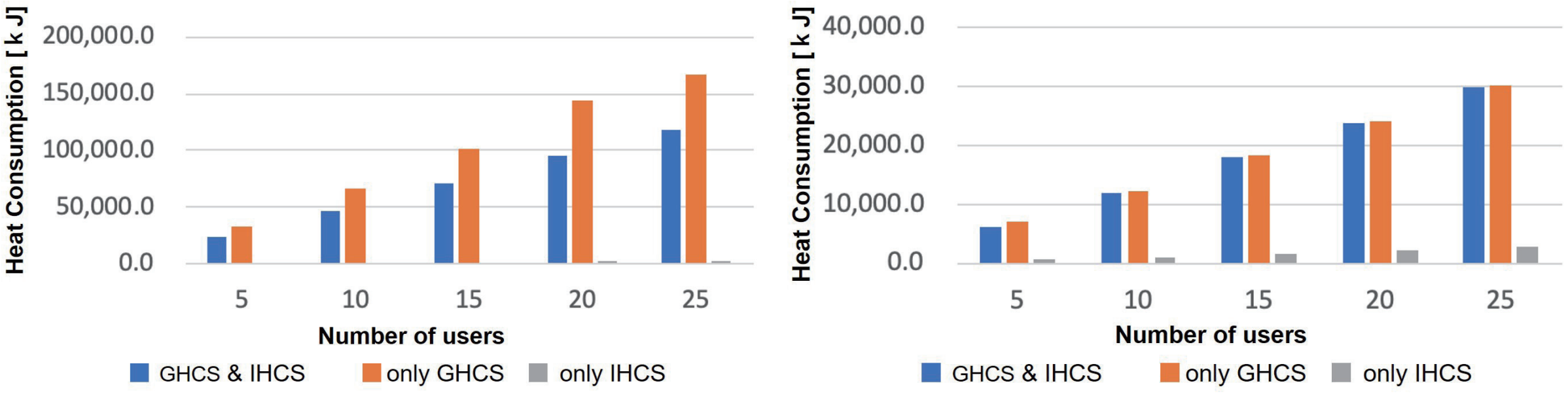

(a) (b)

Fig. 5. (Color online) THC under (a) cooling and (b) heating conditions of each system as functions of number of users in the room.

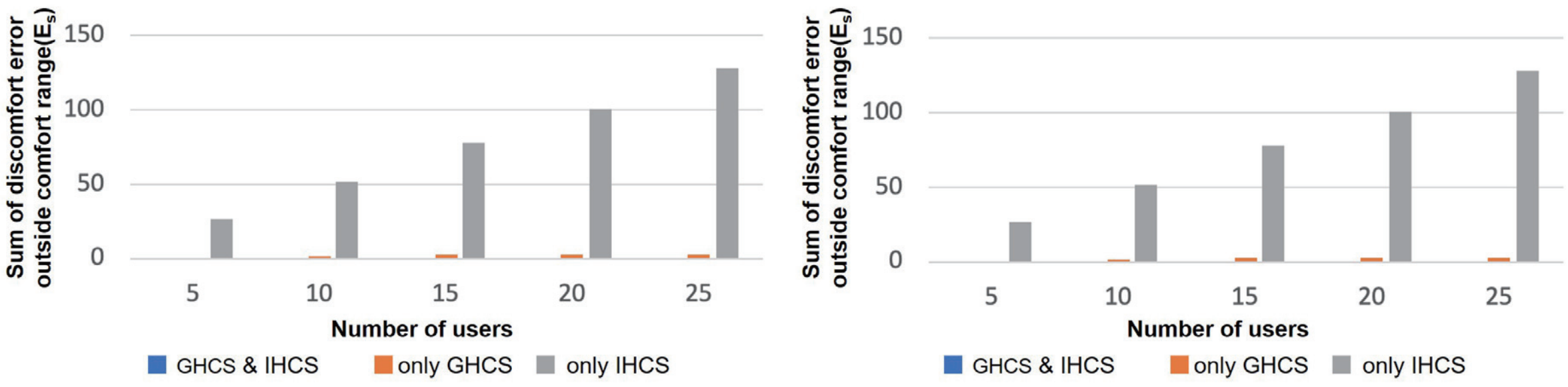

(a) (b)

Fig. 6. (Color online) Sum of discomfort errors outside comfort range ($E_s$) under (a) cooling and (b) heating conditions of each system as functions of number of users in the room.

This result suggests that it is possible to realize both comfort improvement and energy saving for all the users in the room by using the whole air conditioning and heating devices and the individual air conditioning and heating devices together.

## 4. Discussion

We found that under cooling conditions, when using a GHCS and an IHCS together, it is possible to improve the comfort sensation compared with using only an IHCS, and to improve both the comfort sensation and the energy saving compared with using only a GHCS. However, under heating conditions, when using a GHCS and an IHCS together, although it is possible to improve the comfort sensation compared with using only an IHCS or only a GHCS, it is not possible to improve energy saving compared with using only a GHCS, except when the number of users in the room is small.

The possible causes of the difference between the cooling and heating conditions are as follows. The changes in temperature and humidity are different during cooling and heating. The effect of heat generated by the user on the surrounding environment is different between cooling and heating. Moreover, strictly speaking, the amount of heat dissipated and that of water vapor released from the human body vary depending on the ambient temperature. Since these factors are not considered in the model used in this study, it is possible that the effect of the user's heat dissipation on the indoor temperature and humidity was greater than that in the simulation during heating. Therefore, by defining the model of the user's heat generation more strictly, a result closer to the actual situation can be obtained.

Figures 5 and 6 show that neither GHCS nor IHCS can simultaneously manage low heat consumption and total thermal comfort. The design priority of GHCS has always been energy saving rather than thermal comfort. The traditional way of handling well-known differences in preferred temperature between people is to aim for a compromise in an "optimal temperature" at which as few persons as possible are dissatisfied. Indeed, the guidelines for thermal comfort follow current international standards (e.g., the ASHRAE standard 55, one of the most commonly used standards used for regulating building design and operation) that expect a mere 80% satisfaction rate.(13) The best practice for thermal comfort provision has been to restrict indoor temperatures within a narrow range of isothermal conditions with little to no air movement to avoid the generation of draft.(30)

To deal with few dissatisfied persons, Fanger has proposed an individualized thermal control, by which the air is kept at the lowest/highest temperature preferred by most of the occupants and provide the dissatisfied persons with additional moderate local heating cooling devices with which they can control to reach their own preferred temperature.(31) However, it has never been implemented, owing to mainly the issue of the necessity to manually regulate frequently such devices.

The latter issue may be tackled by using an autonomously controlled IHCS with the possibility to directly provide moderate local heating or cooling. However, the efficiency of IHCS alone has been demonstrated to be limited to restricted temperature and humidity ranges for both energy saving and thermal comfort provision. Indeed, in its final report, the Advanced Institute of Wearable Environmental Information Networks recommends the use of personal cooling equipment from 27 °C and joint use with GHCS from 31 °C in summer, and the use of personal warming equipment at 20 °C and its use with GHCS at 16 °C in winter.(28)

Our proposition to use IHCS in combination with GHCS tackles in an efficient way Fanger's design concept and the issues of IHCS single use, since the action of each can be optimized automatically considering all individuals' thermal comfort sensation. Besides, it is also promising considering the fact that only certain parts of the body (e.g., head, wrists, and feet) are crucial for thermal comfort. For example, under uniform environmental conditions, when it is cold, a person's feet and hands feel colder than the other parts of the body, whereas the head is not affected. However, the head is usually more sensitive than the rest of the body to a hot environment.

## 5. Conclusions

In the investigation of a control method that can improve both thermal comfort and energy saving for all users in a coworking space, we proposed a model of thermal comfort of the user and power consumption based on SDI and heat consumption. From the relationship between the thermal comfort sensation of the user and heat consumption depending on the TDI of the room, we established a method of deriving the indoor TDI to achieve both thermal comfort and energy saving. Various environments were simulated with different room sizes, numbers of users in a room, and heating/cooling conditions. The simulation results demonstrated that the proposed solution enables 30 and 10% maximum power consumption reductions compared with using only GHCS under cooling and heating conditions respectively, while keeping the environment in a comfortable range for all users, which is not possible when using only GHCS or IHCS.

## About the Authors

**Guillaume Lopez** received his M.E. degree in computer engineering from INSA Lyon, France, and his M.Sc. and Ph.D. degrees in environmental studies from the University of Tokyo, Japan in 2000, 2002, and 2005, respectively. From 2005, he worked as a research engineer at Nissan Motor Corp. and as a project-dedicated Assistant Professor at the University of Tokyo from March 2009. In April 2013, he joined Aoyama Gakuin University as an Associate Professor of the Department of Integrated Information Technology. His research interests include lifestyle enhancement and healthcare support based on intelligent information systems using wearable sensing technology. His professional memberships include the ACM, AHI, IEEE, IPSJ, and SICE. (guillaume@it.aoyama.ac.jp)

**Takuya Aoki** received his B.Sc. (2017) and M.E. (2019) degrees in information technology from Aoyama Gakuin University, Japan. He has been working at NTT Data Corporation since April 2019. His research interests are in information systems, web services, and the internet of things. (taoki@wil-aoyama.jp)

**Kizito Nkurikiyeyezu** is a Ph.D. candidate at Aoyama Gakuin University, Japan. He received his B.Sc. degree in electrical engineering and M.Sc. degree in electrical and computer engineering from Oklahoma Christian University, Oklahoma City, U.S. His doctoral research takes a multidisciplinary approach and he investigates the possibility of estimating a person's thermal comfort level from the fluctuations in his physiological signals and using appropriate constrained optimization algorithms to provide an optimum and personalized thermal comfort using the least possible energy. (kizito@wil-aoyama.jp)

**Anna Yokokubo** received her M.S. degree (2012) in HCI human-computer interaction (HCI) from Ochanomizu University, Tokyo, Japan. From 2012 to 2017, she worked at Canon Inc., where she contributed to the development of healthcare technology. She is currently working as a research assistant at Aoyama Gakuin University, Kanagawa, Japan. Her research interests are in human computer interaction, information design, and user experience design.
(yokokubo@it.aoyama.ac.jp)